\documentclass[12pt]{amsart}
\usepackage{graphicx}
\usepackage{amssymb}
\usepackage{mathtools}
\usepackage{epstopdf}
\usepackage{comment}
\usepackage{hyperref}

\usepackage{color}
\usepackage{mathpazo}
\calclayout

\usepackage{tikz}
\usepackage{tikzsymbols}
\usepackage[utf8]{inputenc}
\usepackage{times}
\usepackage{mathtools}
\usepackage{footnote}
\makesavenoteenv{table}
\usetikzlibrary{arrows,shapes}
\usetikzlibrary{matrix}
\usetikzlibrary{shadows}
\usetikzlibrary{positioning}

\tikzset{>=stealth',
    pil/.style={
           ->,
           thick,
           shorten <=2pt,
           shorten >=2pt,}
}
\begin{document}

\begin{center}

\vspace{1.5cm}
\huge{\bf Contemplating the fate of modified gravity}
\vspace{1cm}

\large{\bf Alexey Golovnev$^{\dagger}$ and Mar\'ia-Jos\'e Guzm\'an$^*$}  \\
\vspace{1cm}
\end{center}
{\tiny{ $^{\dagger}$Centre for Theoretical Physics, The British University in Egypt, 11837 El Sherouk City, Cairo, Egypt\\ $^*$Laboratory of Theoretical Physics, Institute of Physics, University of Tartu, W. Ostwaldi 1, Tartu 50411, Estonia\\}}
\vspace{1cm}
\begin{center}
{\large{\bf Abstract}}
\end{center}
\small{Possible models of modified gravity are being extensively studied now, with most phenomenological motivations coming from puzzles and tensions in cosmology due to a natural desire to better fit the known and newly coming data. At the same time, available experimental evidence is limited for testing gravity as a force beyond the regimes in which the theory of general relativity has proven to be successful. This situation leads researchers to look for “the simplest modification” to general relativity in a certain class of models, which is enough to solve one or more problems. As a result, we are lost amid a variety of theories with no deeper guiding principle. We give a general review of existing approaches and discuss the current state of the art.}\vspace{1.5cm}
\newline
$^{\dagger}$ email: alexey.golovnev@bue.edu.eg\\
$^*$ email: maria.jose.guzman@ut.ee\\

\normalsize

\section{Introduction} Gravitation is probably the most noticeable and evident physical interaction for every person living in this world. Our current theoretical description of it is known as general relativity (GR), in~which gravity is treated in terms of geometry by~means of a metric $g_{\mu\nu}$  that measures distances in a four-dimensional spacetime characterized by a Lorentzian signature. An~additional important element of a geometric nature is the notion of parallel transport, which is associated with the concept of a connection that is a priori independent of the metric of spacetime but~that in the standard GR is taken to be the usual (pseudo-) Riemannian object of the Levi-Civita connection $\Gamma^{\alpha}{}_{\mu\nu}$.

As is customary in fundamental theoretical physics, theories are defined in terms of Lagrangian mechanics, so that the equations of motion are given by the principle of stationary action. The~main foundational novelty brought by gravity is that we explicitly introduce the metric into the matter action in~order to make it live in a curved spacetime. An~additional feature to take into account is related to fermions, which require the introduction of orthonormal frame fields $e^{a}{}_{\mu}$ known as tetrads or vierbeins via the relation $g_{\mu\nu}=\eta_{ab} e^{a}{}_{\mu} e^{b}{}_{\nu}$ to the Minskowski metric $\eta_{ab}$. 

Gravity itself obtains its dynamics from the Einstein--Hilbert action $\int d^4 x \sqrt{-g}\cdot R$, where $g$ is the determinant of the metric, while $R = g^{\mu\nu}R^{\rho}_{\hphantom{\rho}\mu\rho\nu}$ is the scalar curvature, a~scalar built up from the Riemann tensor $R^{\rho}_{\hphantom{\rho}\mu\rho\nu}$, which describes the way spacetime curves by measuring the change of a vector after it has been parallel-transported along a closed loop~\cite{Misner:1973prb}.

The approach in which we treat the connection coefficients as independent quantities is known as the Palatini formulation. However, the~standard procedure is to directly take the curvature of the Levi-Civita connection, which makes the entire action depend on the metric components only, but~with both the first and second derivatives of the metric entering the Lagrangian. Having higher-than-first derivatives in the action is usually problematic, but~in GR, the second derivatives enter only as a surface term, which does not influence the traditional variational procedure.\footnote{If, as~for example in path integral approaches to quantum theory, the~boundary terms become important, one can find the standard way-out in the Refs. \cite{York:1972sj,Gibbons:1976ue}.} Generally, this blessing is lost upon modifications of the theory, and~then it appears rather hard to meaningfully modify~GR.

At present, we find numerous problems with applying general relativity to the physical world, ranging from issues of quantizing gravity to phenomenological tensions in modern cosmology~\cite{Bull2016,Bernal2016}. Therefore, the~challenging task of modifying gravity acquires not only a theoretical inspiration but also a practical use for current research. Although~there are almost trivial ways to modify a given action, like adding nonlinearity in $R$ corrections as in $f(R)$ gravity, the~mathematical formalism also offers options of more profoundly changing the geometrical~foundations.

Even with the conventional choice of the Levi-Civita connection, one can construct more general scalar invariants in terms of the curvature tensor, or~one can also go for radically changing the connection. The~Levi-Civita connection is entirely determined by requiring it to be symmetric in its lower indices and to have the metric be covariantly constant. If~a connection has an antisymmetric part, we have an additional quantity called torsion, whose effect is to make two covariant derivatives not commute with each other even if acting on a scalar. If~the metric is not covariantly conserved, we are in the presence of a connection with nontrivial nonmetricity, which has the effect of changing the length of a vector upon a parallel transport~\cite{BeltranJimenez:2019esp}.

After reviewing these general definitions, we proceed to discuss the current status of our attempts to modify gravity~\cite{Heisenberg:2018vsk}. Our objective is to give a short but general overview, which should be of interest to anyone who desires to know about the developments in this field of research. In~this essay, we pose a rather philosophical question: do those numerous attempts to modify gravity follow any sound guiding principle? Note that without~a clear understanding of the geometry behind it, even the way to general relativity itself was not smooth, which is captured in the odd hesitations in one of the classical papers~\cite{EinsteinGrossmann}.

\section{Beauty and simplicity} The ultimate goal of theoretical physics is to describe the observed nature in terms of a few simple mathematical principles and their logical consequences. The~criterion for its success or failure is a meticulous comparison with experimental or observational data. Progress is achieved by collaborative efforts of large groups of experts working in diverse areas, ranging from pure mathematics used in physics to engineering solutions for~experiments.
 
The mathematical side of theoretical physics research is now being developed at an astounding pace. Among~the instruments in the physicist's tool kit are differential geometry, functional analysis, abstract algebra, and topology and,~in certain disciplines, utterly abstract tools such as algebraic geometry and number theory. At~the same time, fundamental physics research is driven towards realms which are hardly accessible by experiments, such as enormous, almost-Planck-scale-reaching energies for unified field theories of particle physics or~a whole universe of which we have only one incomplete sample in all our cosmological~observations. 
 
This constitutes a crisis in physics. In~the wetlands of our search, we have a beautifully constructed building named string theory, a~real mathematical skyscraper that rises from its basement made of an extended-objects idea and two-dimensional conformal field theory through all the wonderful floors of algebraic topology and geometry and~up to an invisible top in the clouds that is supposed to yield to us a real-world model via, for~instance, a~proper construction of intersecting stacks of D-branes. The~well-known catastrophe is that there are countless possible worlds there, even under the rather comfortable lamp of largely preserved supersymmetry in the compactified space, however remote it may be from reality, and~with no experimental guidance to look~for.
 
That plight prompted scientists to often judge their theoretical models by considerations of beauty, the~stunning and remarkable mathematical beauty that a lot of our models exhibit. We acknowledge that this approach gives many interesting results and often allows us to better understand the structure and working of our old models when their new modifications are invented or~even to contribute to purely mathematical knowledge. However, in~particle physics, it has also predicted a myriad of unified forces and particles that have not been found, and~the final judge for testing the basic ideas is often expected to lie in the realm of inaccessible energies near the Planck scale. This inevitably digs a deep and yawning gap between us (theoreticians) and experimentalists and,~therefore, the information available about the real world. If~our aim is to describe nature, we do not have to let beauty alone lead us astray as the sole motivating principle to pronounce a theory correct, so that measurements and observations are not only needed but crucial~\cite{Hossenfelder}.
 
Many research goals in gravitational physics might be viewed as less intimidating endeavors. Our common aim is just to find an effective field theory that supersedes general relativity, mostly by how it corresponds to the cosmological data~\cite{CANTATA:2021ktz}. This is not about any abundance of direct experiments, though, and~even considerations of mathematical beauty are much less visible. Nonetheless, we still seek some guiding principle that would allow us to minimally alter the Einstein--Hilbert action in order to obtain alternative extensions. It can be seen as ``simplicity'', a~concept which a has close resemblance to ``beauty'' but~with less mathematical~complexity.

An historical aspect related to the derivation of Einstein's equations is that Einstein struggled for some years to find the correct tensor that would appear on the left-hand side of the field equations. In~a paper from 1913 published with Grossmann as coauthor~\cite{EinsteinGrossmann}, he suggested that the Ricci tensor could play such a role. However, for~the equations in the form of $R_{\mu\nu}= k T_{\mu\nu}$, the~Newtonian limit was not recovered, and this proposal was rejected, leaving the authors to wonder if they should abandon the full general covariance principle. It was only in 1915 that Einstein realized that he could still have a generally covariant equation, and~he wrote the correct field equations and~derived the anomalous precession of Mercury from them. This was the crucial evidence that convinced them of the validity of their theory. Throughout this period, Einstein was driven by the desire to explain this anomaly, which he thought required a new theory of gravity. He was almost alone in this belief, as~most people at that time accepted Seeliger’s hypothesis of the existence of a mass ring around the Sun as a possible explanation~\cite{Woltjer}.
 
The purpose of our current paper is to argue that nowadays, the situation in modified gravity research resembles the old story of Einstein's struggle, i.e.,~there is a lack of a good guiding principle. The~main criterion for modifying gravity is the principle of simplicity, and~unlike real mathematical beauty, this might be even more problematic even though we have not yet come to a common feeling of a severe crisis. The~motivating principles for modifying gravity are no worse than those in a similar search in the standard model of particle physics, and from cosmology, we also have a mild phenomenological push for that, even with only one incomplete sample of a universe, which is seen mostly in terms of the mismatch in different measurements of the Hubble constant. Nonetheless, we are also short of real experimental options, and we do not have such beauty as in several other realms of high-energy~physics.
 
Roughly, the~present-day Hubble ``constant'' ($H_0$) and other related issues are rather hard tensions between direct measurements in the present-day universe and predictions from what we see in the cosmic microwave background (CMB). We possess such data from different coexisting measurements, so that it does not seem plausible that they could all be attributed to some serious fault in astronomical observations.\footnote{We do not consider another possible solution which assumes that the cosmological principle might be wrong~\cite{Banerjee:2020xcn,Lee:2022cyh}.} Given this, such measurements signify either a tremendous change in the well-known physics of matter under well-known, well-studied conditions or a~modification of gravity. In~other words, the~phenomenological justification for current attempts to modify gravity is utterly convincing, but~it does not make these modifications an easy~task.
 
This observational basis is quite far from any intuitive view of gravity as a phenomenon effectively observed as a force or~directly related to an interaction effect between several bodies. We end up modifying gravity in order to change the expansion law of a homogeneous and isotropic universe, a~task which, despite being fundamentally clean and trivial, is the furthest from normal physical intuition. Moreover, we should not wreck the already observed intermediate cosmological times, most importantly in the baryon acoustic oscillations, or~considerably change conditions during the primordial nucleosynthesis, among~others. No undoubtedly successful resolution of these tensions by modified gravity is known to date~\cite{DiValentino:2021izs}, and~with our wording of ``undoubtedly successful'', we simply mean to put this failure in as optimistic a manner as~possible. 
 
Nonetheless, there is also a clear theoretical motivation for this line of research. It is considerably difficult to modify the description of gravity given by the theory of general relativity without tearing down its theoretical viability with the appearance of ghosts or other sorts of instabilities by~modifying the predictions of classical tests such as incorrect bending of light or~by the impossibility of cosmological solutions, for~instance. Therefore, when available options are limited, it seems reasonable to study what can be achieved and how, and~even if we fail in an attempt to build a consistent modification, we do obtain important new angles from which to look at the theory proposed by Einstein. However, in our opinion, a~big problem with this procedure is the lack of mathematical beauty or other reasonable principles guiding us in this~quest.
 
\section{The gravitational swampland} Before we turn to discuss the multiple directions that modified gravity has taken, we should mention that there is still a mild beauty argument in the field, which can also be regarded as stemming from simplicity. Namely, we usually begin from a geometrical picture that could be taken as mathematical beauty~\cite{Goenner:2004se}, but~from the viewpoint of simplicity, it implies that we do not need to construct the entire mechanism of gravity from~scratch. 

A well-known exception is MOdified Newtonian Dynamics (MOND) \cite{Milgrom}. Nonrelativistically, it can be formulated as a nonlinear modification of the Poisson equation \cite{Milgrom2}, which is already not so simple to handle~\cite{meMOND}. Moreover, we ultimately need a relativistic theory, and~then we are thrown back to geometrical concepts. This becomes even more complicated by not immediately producing enough light bending if the extra force acting upon the stars is constructed by using a scalar field with a noncanonical kinetic term; therefore, tensor--vector--scalar models appear~\cite{Bekenstein:2004ne}. In~favor of these models, we can say that it is a rare case of modified gravity which comes from observations of interactions among physical bodies, but~at the same time, it shows that it is not enough, and~cosmology also~matters.
 
\subsection{The smallest amendments} Coming back to the simplest possible generalizations of GR, the~obvious options to be utilized are a cosmological constant and extra dimensions. The~cosmological constant is nowadays not regarded as a modification of gravity, and instead of being a victory, it is actually a rather severe naturalness problem, maybe the most severe problem in all the history of physics from the quantum field theory viewpoint~\cite{Weinberg}. This technical naturalness issue can be thought of as more of an aesthetic (or even psychological) type. One can simply claim that somehow, the renormalized cosmological constant is the way it is in the observations, and~then we can perform all the calculations with no~obstacle. 
 
On the other hand, with~some basic understanding about quantum field theory, we normally expect the dark energy density of order $M^4$, with $M$ being the Planck mass, or a SUSY-violation scale, which is in any case by many orders of magnitude different from its observed scale given rather by the neutrino mass range. This means that the bare cosmological constant should be fine-tuned to the quantum corrections exquisitely and~therefore even depending on the matter content of the theory. To~put it mildly, we do not know the physical origin of such cancellation. There are different opinions about this issue, but~in any case, our point here is that a cosmological constant is not among the main topics of current modified gravity~research.
 
Extra dimensions do not considerably change the picture at the level of the fundamental structure of the theory. However, they constitute a substantial step of introducing new unobserved essences, with~practically no strong motivations if not in string theories or resorting to the argument of simplicity (the simplest way of performing modifications). The~next question is how to hide them, with~one way of proceeding that of placing the observed universe on a brane. This was actually used for degravitation, i.e.,~in an attempt to solve the cosmological constant problem by making large-scale sources unimportant. The~problem is that a codimension one brane was not enough, and~higher codimensional branes are problematic in terms of ghosts. Therefore, this solution ended up introducing an even more complicated structure of cascading gravity which reduces dimensionality in a number of steps~\cite{cascade}. 
 
In this sense, compactified extra dimensions look more appealing, but~then new issues show up, as,~for instance, the impossibility of explaining the accelerated expansion of the visible universe without big changes of its effective fundamental constants due to the unstable size of the extra dimensions unless~serious violations of the energy conditions ($p < -\rho$) are introduced~\cite{nogo}. Despite the fact that there are some caveats~\cite{nonogo} in this claim, it leaves these approaches of almost no modifications largely~unsuccessful.
 
\subsection{Uses of the standard geometry} Having said that, we will turn to the geometric foundations of gravity. The~standard Einstein--Hilbert action is given in terms of the scalar curvature, which is the lowest-order curvature invariant. Therefore, a~logical way to proceed is to consider higher curvature models, i.e.,~keeping up with the same geometric foundations as the standard GR. Many of these generalizations are afflicted (or blessed) by degeneration with scalar--tensor theories, in~the sense that they can be equivalently rewritten as GR with an additional scalar~field.
 
Perhaps the most renowned one of these models is $f(R)$ gravity, which includes the case of Starobinsky inflation~\cite{Star}. It is often claimed to be one of the simplest models, and it is~also a natural choice from the point of view of quantum gravitational corrections. However, in~this case, it is fair to ask why only the scalar curvature is included there, since an honest incorporation of quantum corrections would also require more troublesome terms in the action. Moreover, the~original article by Starobinsky~\cite{Star} is in fact an interesting example of motivations in gravity research, since the original aim was simply to avoid the initial singularity, and~it was not about a physically motivated model of inflation as it is often regarded today. The~ avoidance of the singularity was achieved simply by choosing the one particular trajectory which asymptotes to an exact de Sitter spacetime in the past and~therefore can be considered as practically having no success in resolving~singularities.
 
\subsection{The classes of variation} The aforementioned example was about higher curvature metric theories and~that their main problems are in the higher-derivative-order nature, which normally leads either to ghosts unless in $f(R)$ or~to a frenzy banquet of nonlocal models, where terms with infinitely many derivatives are treated in a special way. Another possible approach is the Palatini formulation, where the connection is assumed to be independent of the metric and usually needs to be assumed symmetric in order to avoid an undesirable projective symmetry. In~this case, some potential problems with the algebraic nature of the equation involving the connection and the matter source were noticed long ago~\cite{probl}. These problems do not seem to be truly fatal, but~the corresponding discussions in the literature about the Palatini formulation~\cite{Olmo} do not look conclusive either. Nevertheless, these models are heavily used for cosmology and astrophysics, and~moreover,  more complicated models of hybrid gravity have been introduced~\cite{hybrid}, which mix both metric and Palatini variational terms in one action. We should ask ourselves, is there indeed enough motivation for that?
 
It is clear that changing the class of variations is one of the methods of modifying a model. For~example, varying in the class of metrics with a fixed value of its determinant gives unimodular gravity. This variation transforms the cosmological constant into a constant of integration, which can be argued to be an improvement of a more psychological rather than empirical value. On~the other hand, if~we take the overall factor in the metric to be not a constant but~rather the canonical kinetic term of an auxiliary scalar field, the~mimetic gravity is obtained~\cite{mimSlava}, receiving its name due to its mimicking of dark matter. There is no doubt that it is intriguing to be able to explain some parts of the dark sector via modifications of gravity. Note, however, that the usual pressureless matter develops caustic singularities in its evolution. These singularities are not catastrophic for any realistically physical model of dark matter due to the breakdown of the pressureless ideal fluid approximation at some point; however, they become much more serious for the mimetic~models.
 
\subsection{New geometries} When the curvature-based options encounter problems, the~next step is to modify the underlying basic geometry~\cite{Hammond:2002rm}. This alternative branch of the modified gravity tree corresponds to such theories which explicitly use connections different from the Levi-Civita one. If~again we summon the elusive doppelg\"{a}nger of simplicity, one straightforward path is to adopt the general metric-affine approach by considering a connection with not only curvature but~also torsion and nonmetricity. Although~a relatively young field of research, the~proliferation of extra degrees of freedom introduced and~the myriad of free parameters to consider in the action destroy the argument of simplicity unless a careful consistency analysis is~made. 
 
Less pretentious candidates are those that uniquely use torsion or nonmetricity as the carrier of gravitation~\cite{Hayashi:1979qx, Aldrovandi:2003pa, Ferraro:2006jd, BeltranJimenez:2019esp}. These models are the so-called (modified) teleparallel or symmetric teleparallel gravity. Among~them, there are equivalents of GR modulo surface terms, i.e.,~at the level of equations of motion. Their dynamics are governed by the torsion scalar $\mathbb T$ or the nonmetricity scalar $\mathbb Q$, respectively. Nevertheless, even the simplest generalizations such as $f(\mathbb T)$ gravity are already radically different from the known curvature-based modified gravities~\cite{Ferraro:2006jd}. Even earlier proposals, including the~so-called new general relativity, are also based on modifications to the teleparallel equivalent of general relativity, although~with a different torsion scalar with free coefficients in front of the Lagrangian terms' quadratic on the torsion tensor~\cite{Hayashi:1979qx}. In~our opinion, these approaches are well-motivated, genuine modifications of GR. When the standard approaches fail, they can teach us something new about gravity~itself. 
 
However, a~big chunk of the literature, past and current, is devoted to applications of this type of model to cosmology. For~the simplest and most basic observables, this is not difficult to do. Indeed, in having a free function at our disposal, we certainly can try to address problems with simple processes such as the expansion law of the universe. Unfortunately, relatively little attention is being paid to foundational issues of $f(\mathbb T)$ and other types of modified teleparallel gravity. At~the same time, the~number of degrees of freedom in the theory is not clear enough and~apparently depends on the background of the tetrads chosen; in~particular, the strong coupling problem is present around the standard cosmological backgrounds~\cite{probWe}. This means that cosmological predictions made in these models are anyhow unreliable since there must be dynamical modes appearing only at nonlinear~orders. 
 
Even before the theoretical intricacies of those models have gained a better understanding, researchers are going for further generalizations. In~addition to using other torsion invariants, higher derivative terms have been introduced, which goes against one of the main initial motivations for modified teleparallel gravity, i.e.,~no higher derivatives in nonlinear modifications. One particular example is $f(\mathbb T,\mathbb B)$ gravity~\cite{Bahamonde:2015zma}, which depends on the torsion scalar $\mathbb T$ and the boundary term $\mathbb B$, which gives the difference between $\mathbb T$ and the Levi-Civita scalar curvature. Evidently, this model can also be presented in terms of $f(\mathbb R, \mathbb T)$ with the Levi-Civita scalar curvature $\mathbb R$, therefore not resolving but~only inheriting the foundational problems of $f(\mathbb T)$ related to the seemingly chaotic nature of local Lorentz violation and remnant~symmetries. 
 
A brief look at the literature shows that models of this mixed torsion--curvature type have been considered in the past~\cite{Myrzakulov:2012axz}, not to be confused with yet another model under the similar name of $f(\mathbb R, T)$ gravity, where $T$ is the trace of the matter energy--momentum tensor~\cite{Harko:2011kv}. However, in order to be able to modify the cosmological evolution, even if having a function $f$ which is linear in both of its variables, another generalization was added to that: shifting the $\mathbb T$ (torsion) and $\mathbb R$ scalars by some arbitrarily chosen functions of the metric and the tetrad components, up~to some specified number of their derivatives (analogous to what is there in the initial scalars themselves), and~for the cosmological applications, those functions are simply reduced to functions of the scalar factor and its derivatives~\cite{Myrzakulov:2012ug}.
 
Needless to say, this approach to modified gravity extends beyond any reasonable motivation to think about modified gravity, and~practically any action in terms of a tetrad can be attributed to this class with~a linear function $f$ by~just subtracting some linear combination of $\mathbb R$ and $\mathbb T$ from the Lagrangian and distributing the result between the two arbitrary new functions. Nevertheless, there is some activity in the research community that is concerned, for~instance, with~applying dynamical system analysis to the cosmology (i.e., one ODE for the scale factor) of some incarnations of this theory, which is chosen by hand at the level of arbitrary functions of the scalar factor and its time derivative~\cite{Papagiannopoulos:2022ohv}. Does it really teach us something new and interesting about gravitation? 
 
\subsection{New fields} Finally, given that the simplest $f(\mathbb R)$ gravity can be represented as a scalar--tensor model, we can also modify gravity by adding new fields, with~the only difference from just some extra component of the matter sector being that these fields should uniformly interact with the usual matter. A~beautiful example of this is represented by supergravity and other ideas coming from string theory. Alternatively, a~popular modern route is to search for the most general second-order derivative theories (Horndeski), then to examine what can be related to such theories by some change in variables (beyond Horndeski); then we can move further and add constraints on the new potentially problematic fields by hand, as~it was performed in a description of mimetic gravity where the scalar field enters only via a constraint~\cite{Golovnev:2013jxa}, or~simply remove some ghosts from the spectrum of previously introduced ill-behaved fields. All of this can be performed for both scalar and vector fields, producing a vast amount of possible models with no clear guiding principle. One prescription that one can find in the literature to classify and propose modifications to gravity is to follow a particle physics perspective of gravity and represent GR as the unique theory of a massless spin-2 particle. By~imposing locality, unitarity, and Lorentz invariance as fundamental principles, the construction of alternative theories emerge naturally based on scalar, vector, and tensor fields~\cite{Heisenberg:2018vsk}. This approach, however, cannot include the geometrical modifications of gravity discussed in this~review.
 
Adding new fields is also a common method of generalizing other models, like $f(\mathbb T,\mathbb B,\phi)$ models in the teleparallel realm, but~with much more numerous examples in the literature, as,~for instance, massive gravity. This model was popular several years ago after it was found as a way to get rid of the Boulware--Deser ghost at all orders~\cite{deRham:2010kj}. However, it appeared to be impossible to construct its cosmological solutions, and~this fact led to generalizations both in the direction of a full-fledged bimetric theory and of models with, for~example, the~graviton mass depending on a new auxiliary scalar field. At~the same time, deep mathematical subtleties related to square roots of matrices~\cite{Golovnev:2017lqm} were never~popular.
 
\tikzstyle{ba} = [rectangle, draw,
text width=4.5em, text centered, rounded corners, minimum height=4em]
\tikzstyle{bb} = [rectangle, draw,
text width=5.5em, text centered, rounded corners, minimum height=4em]
\tikzstyle{line} = [draw, -latex']

\begin{figure}[htbp!]
    \begin{tikzpicture}[node distance = -0.5cm, auto]
    {\huge 
    \node (S) {$S=$};
    \node [right=of S] (cte) {$\dfrac{1}{G}$};
    \node [right=of cte] (int) {$\int$};
    \node [right=of int] (dim) {$d^4 x$};
    \node [right=of dim] (sqr) {$\sqrt{-g}$};
    \node [right=-0.1cm of sqr] (Ric) {$\mathbb R$};
    \node [right=-0.4cm of Ric] (par1) {$($};    
    \node [right=0.1cm of par1] (Gam) {$\Gamma,$};
    \node [right=0.3cm of Gam] (ge) {$g$};
    \node [right=0.1cm of ge] (par2) {$)$};
    \node [right=0.1cm of par2] (pl) {$+$};
    \node [right=0.3cm of pl] (Sm) {$S_M$};
    }
    {\tiny 
    \node [ba, below =1.5cm of cte] (tcte) {variable G};
    \node [below =1.5cm of Ric, outer sep=-0.2pt, inner sep=-0.2pt] (dot) {};
    \node [bb, below =2cm of Ric] (fR) {$f(\mathbb R)$, $f(R_{\mu\nu}R^{\mu\nu})$, ...};
    \node [ba, below =2cm of ge] (mg) {massive gravity};
    \node [below =1.5cm of Gam, outer sep=-0.2pt, inner sep=-0.2pt] (dot2) {};
    \node [bb, below =2cm of Gam] (fT) {Palatini, Metric-affine, $f(\mathbb T)$, $f(\mathbb Q)$};
    \node [bb, below =2cm of dim] (exD) {extra dimensions, non-commuta- tive~spacetime};
    \node [ba, below=2cm of sqr] (mim) {unimodular gravity, mimetic gravity};
    \node [ba, below=2cm of pl] (mass) {bimetric gravity};
    \node [below =1.5cm of pl, outer sep=-0.2pt, inner sep=-0.2pt] (dot3) {};
    \node [bb, below=2cm of Sm] (matt) {scalar- vector- tensor, Horndeski- galileon, ...};
    }
    \draw[->] (tcte) -- node  {{}} (cte);
    \draw[-] (fR) -- node {{}} (dot);
    \draw[->] (dot) -- node {{}} (Sm);
    \draw[->] (dot) -- node {{}} (Ric);
    \draw[-] (fT) -- node {{}} (dot2);
    \draw[->] (dot2) -- node {{}} (Ric);
    \draw[->] (dot2) -- node {{}} (Gam);
    \draw[->] (exD) -- node {{}} (dim);
    \draw[->] (mim) -- node {{}} (sqr);
    \draw[-] (mass) -- node {{}} (dot3);
    \draw[->] (dot3) -- node {{}} (Sm);
    \draw[->] (dot3) -- node {{}} (ge);
    \draw[->] (mg) -- node {{}} (ge);
    \draw[->] (matt) -- node {{}} (Sm);
    \end{tikzpicture}
    \caption{Depiction of modifications to the Einstein-Hilbert Lagrangian. The numeric factors in front of the action have been absorbed in the constant $G$.}
    \label{la_figura}
\end{figure}
 
\subsection{The general picture} Considering that the reader might be overwhelmed by the multitude of extensions and generalizations of general relativity, we aim to summarize our point and make an attempt to schematize such a variety of modifications as changes of a particular element of the Einstein--Hilbert Lagrangian. As~shown in Figure~\ref{la_figura}, some modifications can be associated with more than one element of the Lagrangian. This indicates that for some alternative theories, for~instance such as $f(\mathbb R)$ gravity, the~nonlinear deformations can be mathematically (and possibly physically) interpreted as a matter content. Some analogous diagrams occasionally appear in the literature~\cite{CANTATA:2021ktz,Baker}. Although~sometimes inaccurate or misleading, they can be regarded as faithful attempts to expose the rich variety of theories, but they lack  
 a genuine effort to understand the physics behind them. Certainly, our scheme in Figure~\ref{la_figura} is highly incomplete and biased towards our research interests and limited knowledge, and~it does not include some more radical generalizations, like Finsler geometry, which are hard to relate to some particular part of the EH~action.
 
In Figure~\ref{la_figura}, each arrow requires some interpretation which is not always directly straightforward. For~example, the~variation-class-changing models (unimodular and mimetic gravity) are related to the $\sqrt{-g}$ measure part, for~they somehow specify the overall factor in the metric. At~the same time, mimetic gravity, as~well as several or almost all other theories, can also be rewritten as an extra part in the matter sector. Some relatively young theories, such as those living in the metric-affine or teleparallel realm ($f(\mathbb T)$,$f(\mathbb Q)$, etc.), change the connection but~also replace the Ricci scalar with an alternative scalar (different from it by a boundary term), and~at the same time, they admit an Einstein-frame-like representation whose identification with a well-studied matter content is not yet well understood~\cite{Golovnev:2019kcf,BeltranJimenez:2021auj}.

\section{Quantum gravity?} Note that we did not mention the problems of quantum gravity in realms such as string theory, loop quantum gravity, asymptotic safety, or~different approaches available in the literature. In~the opinion of the majority, gravity must finally be of a quantum nature. Nonetheless, this notion has serious problems, not only technical but also conceptual. The~common viewpoint is that we have to modify gravity in~order to make it better suited for quantization. However, actually, quantum field theory is not a mathematically well-defined field in itself. We lack clear mathematical nonperturbative definitions, and axiomatic field theory does not extend beyond general statements like the CPT theorem and constructions of super-renormalizable models such as $1+1$-dimensional scalar fields. These unclear aspects can lead us to think that we still miss some truly important aspect about quantum theory and~not only about~gravity.

Let us impart one additional quantum gravity remark here. The~basic nature of the gravitational interaction is that the energy momentum plays the role of its charge. In~other words, the~higher the energy of the colliding particles, the~stronger they interact. This points at the growth of amplitudes with energy, and as such, it speaks against renormalizability. Then we should either find a radical modification of gravity for quantization (as in the aforementioned spectrum of theories) or~we need a fully nonperturbative understanding of quantum physics. Of~course, we can say that classicalization ideas~\cite{Dvali:2010jz} point towards this direction, but~this direction 
 then resembles more an attitude of closing the eyes when it comes to asking about what happens behind the newly formed~horizon.
  
\section{Conclusions} We acknowledge a sense of being lost in a labyrinth of modified gravity models. A~meaningful modification of GR is not straightforward to find, but~possible ways to go are ubiquitous. The~experimental determination of the velocity of gravitational waves, though~restrictive~\cite{gravconstr}, is not enough to radically reduce the space of options. A~rather common motivation in most research articles is to take the simplest possible modification of gravity in a given class. The~simplicity principle is basically reduced to ``we try out whatever we can, and~watch what comes out''. In~this line, it is also even more worrisome to generalize already generalized versions of GR and~to consider this approach useful just because the more generalized theory contains an already known, but~untested, theory of~gravity. 

Oftentimes, we see the appearance of studies of linear cosmological perturbations in models with a guaranteed severe strong coupling problem or~works constructing some models by hand with no deeper meaning behind them and then applying beautiful mathematical techniques to them, like dynamical systems for cosmology. These approaches wield an attitude of calculating more and thinking less, which is for sure a kind of forced approach in the modern ``publish or perish'' research culture. However, what is the future to come?

There is nothing we can do about some empirical facts, like our technological limitations. We can neither build a particle accelerator of the solar system's size nor create a large set of universes with~intelligent life and without. Therefore it is an objective reality that many questions we ask cannot be addressed by the simple ``go and have a look'' prescription. This does not imply that by any means we have to stop our research activity, but~we should rather exercise caution and honesty when it comes to assessing whether we are describing the real world or just studying mathematical properties of possible theories. In~our opinion, both directions are good as long as they are truthfully undertaken and~presented.

We would like to emphasize that, fortunately, both mathematical and phenomenological directions are available and open in the modified gravity fields, but~they should be exercised responsibly. As~we mentioned above, it is probably not of much significance to study predictions of linear cosmological perturbation theory in a gravity model that has a strong coupling or~other features that make its predictions unreliable, though~that is not to say that it should never be computed at all. There might be some vagueness about the derivations of the fundamental issues, some independent theoretical interest of studying this ill-posed particular model, or~a hope that some more general models of this type would be healthier and with the same general features of perturbations. Although~many reasons are possible, we must be clear about what we are doing and~why.

The final point to make here is that we should give more attention to studying the landscape of possible modified gravity models and their classification and physical equivalences. It is highly necessary to have a general picture of possible generalizations and~also the net of equivalences between them, and~last but not least, we ought to study the foundational issues of all these models in order to know which of them can ever be viable. On~the one hand, it might spare lots of effort being wasted in studying hopeless cases in great detail. On~the other hand, it gives us a much better and deeper theoretical understanding of what gravity is and how it may or may not work. We hope that our presentation will help the new generation of modified gravity researchers obtain a panoramic picture of the current state of the art and develop their own opinions on how to move~forward.

{\tiny {\bf Acknowledgments}. MJG has been supported by the Estonian Research Council grant PSG910.}

\end{document}